\documentclass[fleqn,usenatbib]{mnras}

\usepackage{newtxtext,newtxmath}

\usepackage[T1]{fontenc}

\DeclareRobustCommand{\VAN}[3]{#2}
\let\VANthebibliography\thebibliography
\def\thebibliography{\DeclareRobustCommand{\VAN}[3]{##3}\VANthebibliography}

\usepackage{graphicx}	
\usepackage{amsmath}	

\defcitealias{2017Robitaille}{\scshape R17}
\defcitealias{2009Gruendl}{\scshape G09}
\defcitealias{2009Seale}{\scshape S09}

\title[JWST Observations of M33]{JWST Reveals Star Formation Across a Spiral Arm in M33}

\author[J. Peltonen et al.]{
Joshua Peltonen,$^{1}$\thanks{E-mail: peltonen@ualberta.ca}
Erik Rosolowsky,$^{1}$
Thomas G. Williams,$^{2}$
Eric W. Koch,$^{3}$
Andrew Dolphin,$^{4,5}$
\newauthor J\'er\'emy Chastenet,$^{6}$
Julianne J. Dalcanton,$^{7,8}$
Adam Ginsburg,$^{9}$
L. Clifton Johnson,$^{10}$
Adam K. Leroy,$^{11,12}$
\newauthor Theo Richardson,$^{9}$
Karin M. Sandstrom,$^{13}$
Sumit K. Sarbadhicary,$^{11,12,14}$
Adam Smercina,$^{7}$
Tobin Wainer,$^{7}$
\newauthor Benjamin F. Williams$^{7}$
\\
$^{1}$Dept. of Physics, University of Alberta, Edmonton, Alberta, Canada, T6G 2E1\\
$^{2}$Sub-department of Astrophysics, Department of Physics, University of Oxford, Keble Road, Oxford OX1 3RH, UK\\
$^{3}$Center for Astrophysics $\mid$ Harvard \& Smithsonian, 60 Garden St., 02138 Cambridge, MA, USA\\
$^{4}$Raytheon, 1151 E. Hermans Road, Tucson, Arizona 85756, USA\\
$^{5}$Steward Observatory, University of Arizona, 933 N. Cherry Avenue, Tucson, Arizona 85719. USA\\
$^{6}$Sterrenkundig Observatorium, Universiteit Gent, Krijgslaan 281 S9, B-9000 Gent, Belgium\\
$^{7}$Department of Astronomy, University of Washington, Box 351580, Seattle, WA 98195-1580, USA\\
$^{8}$Center for Computational Astrophysics, Flatiron Institute, 162 Fifth Avenue, New York, NY 10010, USA\\
$^{9}$Department of Astronomy, University of Florida, PO Box 112055, USA\\
$^{10}$Center for Interdisciplinary Exploration and Research in Astrophysics (CIERA) and Department of Physics and Astronomy, Northwestern University,\\
1800 Sherman Ave., Evanston, IL 60201, USA\\
$^{11}$Department of Astronomy, The Ohio State University, 140 West 18th Avenue, Columbus, OH 43210, USA\\
$^{12}$Center for Cosmology and Astroparticle Physics, 191 West Woodruff Avenue, Columbus, OH 43210, USA\\
$^{13}$Department of Astronomy \& Astrophysics, University of California, San Diego, 9500 Gilman Drive, San Diego, CA 92093, USA\\
$^{14}$Department of Physics, The Ohio State University, Columbus, OH 43210, USA\\
}

\date{Accepted XXX. Received YYY; in original form ZZZ}

\pubyear{2023}

\begin{document}
\label{firstpage}
\pagerange{\pageref{firstpage}--\pageref{lastpage}}
\maketitle

\begin{abstract}
Young stellar objects (YSOs) are the gold standard for tracing star formation in galaxies but have been unobservable beyond the Milky Way and Magellanic Clouds. But that all changed when the James Webb Space Telescope was launched, which we use to identify YSOs in the Local Group galaxy M33, marking the first time that individual YSOs have been identified at these large distances. We present MIRI imaging mosaics at 5.6 and 21~$\mu$m that cover a significant portion of one of M33's spiral arms that has existing panchromatic imaging from the Hubble Space Telescope and deep ALMA CO measurements. Using these MIRI and Hubble Space Telescope images, we identify point sources using the new DOLPHOT MIRI module. We identify 793 candidate YSOs from cuts based on colour, proximity to giant molecular clouds (GMCs), and visual inspection. Similar to Milky Way GMCs, we find that higher mass GMCs contain more YSOs and YSO emission, which further shows YSOs identify star formation better than most tracers that cannot capture this relationship at cloud scales. We find evidence of enhanced star formation efficiency in the southern spiral arm by comparing the YSOs to the molecular gas mass.
\end{abstract}




\section{Introduction}

Understanding star formation is essential to deciphering galaxy evolution. Typically, star formation has been observed through several multi-wavelength tracers such as H$\alpha$, far-ultraviolet, infrared and radio continuum. These tracers have shown that star formation is intimately linked with molecular gas through the Kennicutt-Schmidt relation \citep{1959Schmidt,1988Kennicutt}. However, this relation breaks down when observing star formation tracers at the scales of individual giant molecular clouds \citep[GMCs;][]{2010Onodera,2010Schruba}. This is likely because many of these tracers, such as H$\alpha$ and UV, capture the formation of massive stars only after they have had sufficient time to ionize and clear the surrounding molecular gas. This delay makes directly comparing star formation tracers to individual GMCs difficult.

The most direct way to measure the star formation process is to observe young stellar objects (YSOs). YSOs allow the star formation process to be detected earlier and more directly than many common extragalactic star formation tracers, encompassing the deeply embedded phase to the later pre-main sequence stage \citep{1987Lada}. YSOs have been studied extensively in nearby Milky Way GMCs \citep[e.g.,][]{2009Evans,2011Williams,2015Stutz}, primarily using infrared observatories (e.g., \textit{Spitzer Space Telescope} and \textit{Herschel Space Observatory}) that can penetrate the dense, dusty gas in which the YSOs are embedded. These studies have shown that for solar neighbourhood GMCs, star formation correlates well with molecular gas \citep{2010Lada}. Observing large populations of YSOs beyond the solar neighbourhood in the Milky Way is challenging because of extinction, line-of-sight confusion, and distance ambiguities \citep[e.g.,][]{2009Roman,2015Eden,2018Motte}. Therefore, understanding how star formation proceeds in different galactic environments requires observations of YSOs in other galaxies.

With the previous generation of telescopes, individual YSOs beyond the Milky Way have only been observed in the Milky Way satellite galaxies. This is due to the relatively poor resolution of previous observatories, which could only capture the high-mass YSOs in the Magellanic clouds. For example, \citet{2008Whitney} and \citet{2009Gruendl} (G09) used \textit{Spitzer} observations to identify massive YSOs in the Large Magellanic Cloud (LMC). To eliminate various IR bright contaminants, \citet{2008Whitney} used several different colour cuts to isolate YSOs. Alternatively, \citetalias{2009Gruendl} used visual inspection, location information, and a more minimal set of colour cuts to identify YSOs. \citet{2013Sweilo} employed a similar methodology in the Small Magellanic Cloud (SMC) by using a combination of colour cuts and visual inspection to identify YSOs. Follow-up spectroscopy from \citet{2009Seale} (S09) in the LMC and \citet{2013Oliveira} in the SMC showed that colour cuts with visual inspection effectively identify YSOs with few contaminants. \citet{kinson2022} applied machine learning to near-infrared surveys of M33 to identify YSOs. However, YSOs have much greater overlap with potential contaminants in the near-infrared.

The \textit{James Webb Space Telescope} \citep[JWST;][]{jwst-mission} has opened the opportunity to find massive YSOs throughout the Local Group \citep[e.g.,][]{2023Lenkic}. An excellent candidate is the Triangulum galaxy (hereafter M33). The moderate inclination of M33 \citep[55$^\circ$; ][]{koch18} allows for minimal extinction and relative positions of objects to be found without ambiguity. At the distance of M33 \citep[859~kpc,][]{2017distanceM33}, massive YSOs (M$\gtrsim$6~M$_\odot$) are readily detectable in the mid-infrared with the Mid-Infrared Instrument \citep[MIRI;][]{miri-paper} onboard JWST. 

M33 is similar in star formation rate, size and metallicity to the LMC but with clearly visible flocculent spiral arms \citep{1980Humphreys,2007Rosolowsky,2018Dobbs,PHATTERII}. These spiral arms offer a chance to measure how the presence of a spiral arm affects star formation. There is certainly more star formation happening in spiral arms, but whether spiral arms are only concentrating or also enhancing star formation remains unclear \citep[e.g.,][]{2010Foyle,2017Leroy,2018Hirota}. These studies of star formation in spiral arms have been conducted using tracers of high-mass star formation. Now, we have the opportunity to address this question of concentration or enhancement directly using YSOs, recognizing that a flocculent spiral like M33 does not represent all spiral structure phenomena. 

Here we present JWST MIRI observations at 5.6~$\mu$m and 21~$\mu$m, covering a large portion of the southern spiral arm in M33. Using these observations along with existing \textit{Hubble Space Telescope} (\textit{HST}) observations from \citet{2021PHATTERI}, we identify YSO candidates using colour selections, positional information, and visual inspection, employing a methodology similar to \citetalias{2009Gruendl}. Candidates can then be used to measure the star formation and efficiency both inside and outside a spiral arm at individual cloud scales.

 \begin{figure*}
	\includegraphics[width=\textwidth]{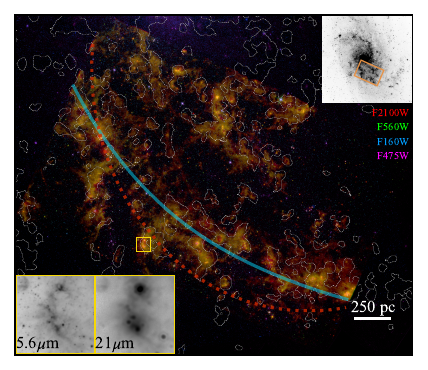}
\caption{A four-colour image showing the MIRI data and \textit{HST} data from the PHATTER survey. The red channel is the MIRI filter F2100W (21~$\mu$m), the green channel is the MIRI filter F560W (5.6~$\mu$m), the blue channel is the \textit{HST} filter F160W (1.6~$\mu$m), and the purple filter is the \textit{HST} filter F475W (0.475~$\mu$m). The white outlines show the boundaries of the GMCs based on $^{12}$CO(2-1) ALMA ACA observations (Koch et al. in preparation). The light blue line shows the logarithmic spiral arm model. The red dotted line shows the \citet{smercina23} spiral arm model, where Section \ref{sec:spiralarm} describes these two spiral arms. The top right shows a \textit{B} band image of M33 from the 4~m Mayall telescope \citep{2006Bband} with the JWST MIRI coverage outlined in orange. Finally, the bottom left panels show the quality of the two MIRI filters in more detail, with the corresponding area in the main figure shown as a yellow box.}
    \label{fig:miridat}
\end{figure*}

In Section \ref{obs}, we first present an initial analysis of the new MIRI observations of M33, followed by a description of the Panchromatic Hubble Andromeda Treasury: Triangulum Extended Region (PHATTER) survey \citep{2021PHATTERI}. Section \ref{obs} also briefly describes the identification of GMCs from ALMA ACA data and the mapping of the spiral arm using atomic ISM traced by the VLA. We then introduce the new DOLPHOT \citep{2000Dolphin,2016Dolphin} MIRI module in Section \ref{sec:dolphot}, which we use to measure photometry of point sources from the MIRI and \textit{HST} data. Section \ref{ysoiden} details the identification of YSO candidates from the DOLPHOT point sources using colour cuts and position information. We then use these YSO candidates in Section \ref{ysoana} to determine the relation between GMCs and the number and flux of YSOs and compare it to Milky Way GMCs. We then find the spiral arms' effect on star formation efficiency. Finally, Section \ref{conc} provides a summary of our findings.

\section{Data}\label{obs}
We use several data sets that have observed the southwest region of M33. For this analysis, we adopt a distance to M33 of 859~kpc \citep{2017distanceM33} and orientation parameters from \citet{koch18}, namely an inclination of $i=55^\circ$ and a kinematic position angle of $\phi_0 = 201^\circ$.

\subsection{JWST Observations}
Under JWST GO program 2128, we observed M33 using the MIRI instrument \citep{miri-paper} onboard JWST as the primary camera operated in imaging mode. We observed a $5\times 5$ tile mosaic in the galaxy that covers a $0.20^\circ \times 0.16^\circ$ region on the sky, which projects to an area of $5.5$ kpc$^2$.  The survey is centred on RA$=23.4376^\circ$ and DEC$=30.5813^\circ$, covering a significant portion of one of M33's spiral arms (Figure \ref{fig:miridat}). Since the emission from the galaxy fills the detector, we also observed an off-galaxy background position located $0.8^\circ$ east of our field centre that was selected to be free of emission from the galaxy. In addition to setting the MIRI background levels, we use this field to estimate the number of background galaxies expected to contaminate the primary science observations. We observed in the 21~$\mu$m (F2100W) and 5.6~$\mu$m (F560W) filters. We also collected data using the NIRCam instrument \citep{NIRCAM} in parallel, which covers the center and northwest portion of the galaxy and will be presented in future work. We observed using a 4-point dither pattern optimized for parallel MIRI and NIRCam observations. We used the FASTR1 readout pattern with 25 groups/integration and 2 integrations, leading to a total integration time of 566 seconds per field.

We reduced the data using the PHANGS-JWST imaging pipeline \citep{2023Lee} with improvements made as JWST Cycle 1 has progressed (Williams et al., in preparation). This pipeline customizes processing using the regular JWST processing pipeline\footnote{\url{https://jwst-pipeline.readthedocs.io/}} and applies post-processing steps to optimize pointing and image quality. The pipeline was used with JWST Calibration Reference Data System context \texttt{jwst\_1077.pmap}. The pipeline produces individual calibrated ``Level 2'' data of each dither position for each field, which are cosmic-ray subtracted and then used for point source photometry. These Level 2 data are combined using the PHANGS-JWST and main JWST pipelines to form a ``Level 3'' image mosaic, which we use to compare with other wavebands. As part of the global astrometric alignment process in Level 3, the pipeline corrects the astrometry of the Level 2 data used for photometry. The estimated noise levels for the Level 2 data are 3.2 $\mu$Jy in F2100W and 0.033 $\mu$Jy in F560W. Figure \ref{fig:miridat} shows the 21~$\mu$m image (red channel) and the 5.6~$\mu$m (green channel).

\subsection{PHATTER}\label{sec:PHATTER}
The MIRI survey area is contained within the PHATTER \textit{HST} survey area \citep{2021PHATTERI}. The PHATTER survey is composed of six \textit{HST} bands from the near-infrared to ultraviolet. The filters F160W and F475W are included in Figure \ref{fig:miridat} (blue and purple channel). Using the PHATTER survey, \citet{smercina23} identified several populations of stars in different evolutionary states. These populations range from main-sequence stars ($\sim$80~Myr) to red giant branch (RGB) stars ($\sim$4~Gyr). \citet{smercina23} found that the older RGB stars do not exhibit the flocculent spiral structure typically observed in M33. Instead, these RGB stars show a barred spiral with two distinct grand design arms. We use the Southern spiral arm model from \citet{smercina23}, shown in Figure \ref{fig:miridat} as a red dotted line.

\subsection{ALMA ACA} 
The MIRI survey area is also covered by a new $^{12}$CO J=2-1 survey (2017.1.00901.S; 2019.1.01182.S) using the Atacama Large Millimeter/submillimeter Array (ALMA) Atacama Compact Array (ACA). The ACA survey has a linear resolution of $\approx35$~pc ($\approx8\farcs5$ beam size) at the distance of M33. The survey and GMC catalogue are presented in more detail in Koch et al. (in preparation), and we provide only a brief description here. 
Koch et al. (in preparation) obtained a catalogue of 444 GMCs using the Spectral Clustering for Molecular Emission Segmentation (SCIMES) algorithm \citep{2015SCIMES}, of which 106 are fully within the MIRI survey area, and 16 are partially covered.
The white contours in Figure \ref{fig:miridat} show the location and extent of the GMCs. This work relies primarily on the position and measurements of the GMC H$_2$ mass. 
The luminous H$_2$ masses were estimated using a metallicity dependent CO-to-H$_2$ conversion factor as described in \citet{sun20} ($\alpha_\mathrm{CO} \propto Z^{-1.6}$) and a constant CO(2-1)/CO(1-0) line ratio of 0.8 from \citet{druard14}.  We use the metallicity gradient from \citet{bresolin11} and a solar metallicity of $Z_\odot = 8.7$ such that $\alpha_\mathrm{CO(2-1)} = [11.4~M_\odot~\mathrm{pc}^{-2} / (\mathrm{K~km~s}^{-1})](R_\mathrm{gal}/\mathrm{kpc})^{0.06}$.

\subsection{VLA 21-cm {\sc HI} Data}\label{spiralarm}

To trace the atomic ISM, we use 21-cm {\sc H\,i} VLA data from \citet{koch21}, which includes short-spacing information from the Green Bank Telescope (GBT).
These data have a spatial resolution of $8''\approx32~\mathrm{pc}$ and were imaged with 1~km/s velocity channels.
We use the \textsc{H\,i} velocity information to define the locus of the Southern spiral arm to compare and contrast with the arm location derived from stellar populations. The \textsc{H\,i}-defined arm is shown as the cyan line in Figure \ref{fig:miridat}.

\section{DOLPHOT} \label{sec:dolphot}
We use the DOLPHOT \citep{2000Dolphin,2016Dolphin} JWST/MIRI module to extract stellar photometry for the point sources simultaneously in the 21, 5.6, and \textit{HST} 1.6 $\mu$m bands. The MIRI module, which can be found on the main DOLPHOT website\footnote{\url{http://americano.dolphinsim.com/dolphot/}}, was developed using many of the same code and methods used for the JWST/NIRCam and JWST/NIRISS modules \citep{2023Weisz}.  Key differences between the MIRI module and the other JWST modules include the following. MIRI-specific PSF libraries and associated encircled energy values were calculated for each filter at 25 locations using WebbPSF \citep{2014Perrin}. As with the other JWST modules, WebbPSF development version 1.0.1 was used with distorted PSF modules, OPD maps from 6/24/2022, 5$\times$ oversampling, and an input G5V model spectrum from the Phoenix stellar library \citep{2013Husser}.  Support for photometry in both the primary imaging field of view and the Lyot coronagraph is enabled by default, though the Lyot region may be masked by the user.  Finally, photometry is available in either ABmag/Jy or VEGAmag as a user-specified parameter to DOLPHOT, with ABmag to VEGAmag conversions obtained from the \texttt{jwst\_miri\_abvegaoffset\_0001} ASDF reference file. This feature has subsequently been added to the NIRCam and NIRISS modules.

We use the recommended settings for FitSky=2 from the DOLPHOT MIRI module manual to identify sources in the MIRI field. We have chosen to mask the Lyot coronagraph because we found the astrometry to be less accurate. The DOLPHOT output gives the flux of the sources in 5.6~$\mu$m ($F_{5.6}$) and 21~$\mu$m ($F_{21}$) and the Vega magnitudes in 1.6~$\mu$m. In order to compare the flux from MIRI to Vega magnitudes, we first convert the Vega magnitudes to AB magnitudes using $M_\mathrm{AB}-M_\mathrm{Vega}=1.274$ for F160W \citep{hubIRdoc}. We then convert AB magnitudes into flux at 1.6 $\mu$m ($F_{1.6}$). Flux uncertainties are estimated based on Poisson statistics applied to count measurements, determined from the count-per-second rates from ramp fitting and then multiplying by the exposure time reported in the header \texttt{EFFEXPTM}.

We only use DOLPHOT sources with object type 1 (stars) and no quality flags in the three bands. We also remove sources with crowd $\geq$ 1, sharpness $< -0.5$, and round $> 0.5$. The MIRI filters contain diffuse emission from warm dust, which DOLPHOT occasionally identifies as point sources. To avoid these ISM detections, we only include sources with a local signal-to-noise ratio (SNR) $>$ 90 in F560W and $>$ 30 in F160W, which was found by eye to eliminate the majority of these ISM sources. These cuts leave 7895 point sources in the survey area. 

We also use DOLPHOT to find sources in the off-galaxy background images as a check on the number of contaminants we expect from background galaxies and Milky Way stars.  Using the same sample selection as for the science field photometry, we find 36 sources.

\section{YSO Identification}\label{ysoiden}

 \begin{figure}
	\includegraphics[width=\columnwidth]{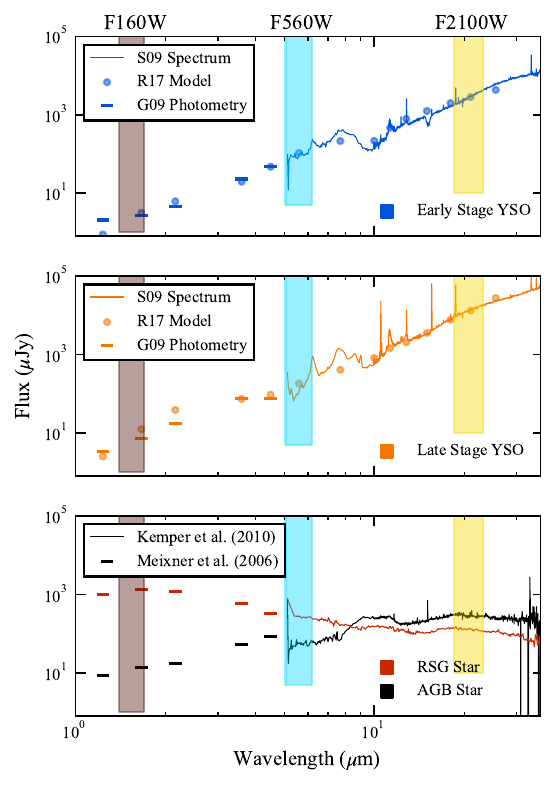}
\caption{Example of mid-infrared spectra and near-infrared photometry of YSOs and contaminants from the LMC that have been shifted to the distance of M33. Shown here are two YSOs (blue and orange) from \citetalias{2009Seale} and \citetalias{2009Gruendl} along with two contaminants. The contaminants are a red supergiant (red) and an asymptotic giant branch star (black) with spectra from \citet{2010Kemper} and photometry from \citet{2006Meixner}. The three shaded regions show our MIRI filter (F560W and F2100W) and the \textit{HST} filter (F160W) choice, with the bottom of the shade indicating the 5$\sigma$ detection limit of that filter. The simulated observations of the two YSOs have been fit to an \citetalias{2017Robitaille} model based only on the three filters. The flux of these model fits from every other filter are also shown as dots. Both the deeply embedded YSO shown in blue and the less embedded YSO shown in orange are well fit with \citetalias{2017Robitaille} models and can be distinguished from the contaminants.}
    \label{fig:simObs}
\end{figure}

\subsection{Filter Selection}

The two MIRI filters (F560W, F2100W) in our survey were chosen to optimize the detection of continuum YSO emission and separate from contaminants, including red giants and background sources, following the guidance from  \citet{jones17}, which emphasizes the importance of selecting at least two continuum-dominated bands to identify YSOs with clear demonstration of the use of F2100W.  We selected F560W as a bluer filter that is continuum-dominated for YSOs instead of the F1000W favoured in \citet{jones17} because of the significantly better spatial resolution, which is essential for minimizing blending and crowding.

To demonstrate the effectiveness of YSO selection from these two filters, we fit YSO models to real YSOs as if they were only observed in these filters. Figure \ref{fig:simObs} shows the photometry and spectroscopy from two observed massive YSOs in the LMC and two non-YSOs from the LMC for comparison. The spectra for the YSOs are from \citetalias{2009Seale}, showing objects in the LMC originally identified as YSOs by \citetalias{2009Gruendl} with broad-band photometry. The common contaminants are a red supergiant and an asymptotic giant branch (AGB) star identified by photometry from \citet{2006Meixner} and spectra from \citet{2010Kemper}. The YSOs show the characteristic spectral energy distribution of a dust-embedded source, rising toward the longer wavelengths, with a prominent silicate absorption feature near 10 $\mu$m. However, the contaminants have spectra that fall or remain nearly constant with wavelength.

We optimize our selection of YSOs for the F160W, F560W, and F2100W filter set using a suite of models from \citet{2017Robitaille} (R17). Specifically, we use model set \texttt{spubsmi} with 40000 models all with a disk, envelope and ambient ISM. Each of these models has nine viewing angles with unique fluxes. However, using the other model sets with disk, envelope, and ambient ISM (\texttt{spu-hmi}, \texttt{spu-smi}, and \texttt{spubhmi}) has little effect on our results. For all of the \texttt{spubsmi} models, we perform simulated observations using the MIRI filters and the \textit{HST} filter F160W. 

Figure \ref{fig:simObs} illustrates that real massive YSOs have colours that can be reproduced by \citetalias{2017Robitaille} models when fit only to the three filters in our set. The \citetalias{2017Robitaille} models match the spectra and photometry of the observed YSOs except for the features at 7.7~$\mu$m (polycyclic aromatic hydrocarbon emission) and 10~$\mu$m (silicate absorption), which vary greatly with evolutionary phase \citepalias{2009Seale} and the surrounding ISM. 

These two MIRI filters and the F160W filter from \textit{HST} are sufficient to separate the typical rising spectra of YSOs from the falling or constant spectra of many contaminants. When combined with positional information, the three filters should be sufficient to eliminate many non-YSO objects. However, the evolutionary state of the YSOs is indistinguishable from these three filters alone. In future work, we will explore how well the physical parameters of individual YSOs can be constrained with these observations, but our primary goal here is to optimize recovering YSO candidates (YSOC) from the available filters.

\subsection{YSO Selection}

We identify YSOs from the DOLPHOT catalogue of point sources with the goal of finding the most complete sample while minimizing contaminants as much as possible. Our approach uses a colour-colour diagram, a colour-magnitude diagram (CMD), proximity to GMCs, and visual inspection. 

\begin{figure}
	\includegraphics[width=\columnwidth]{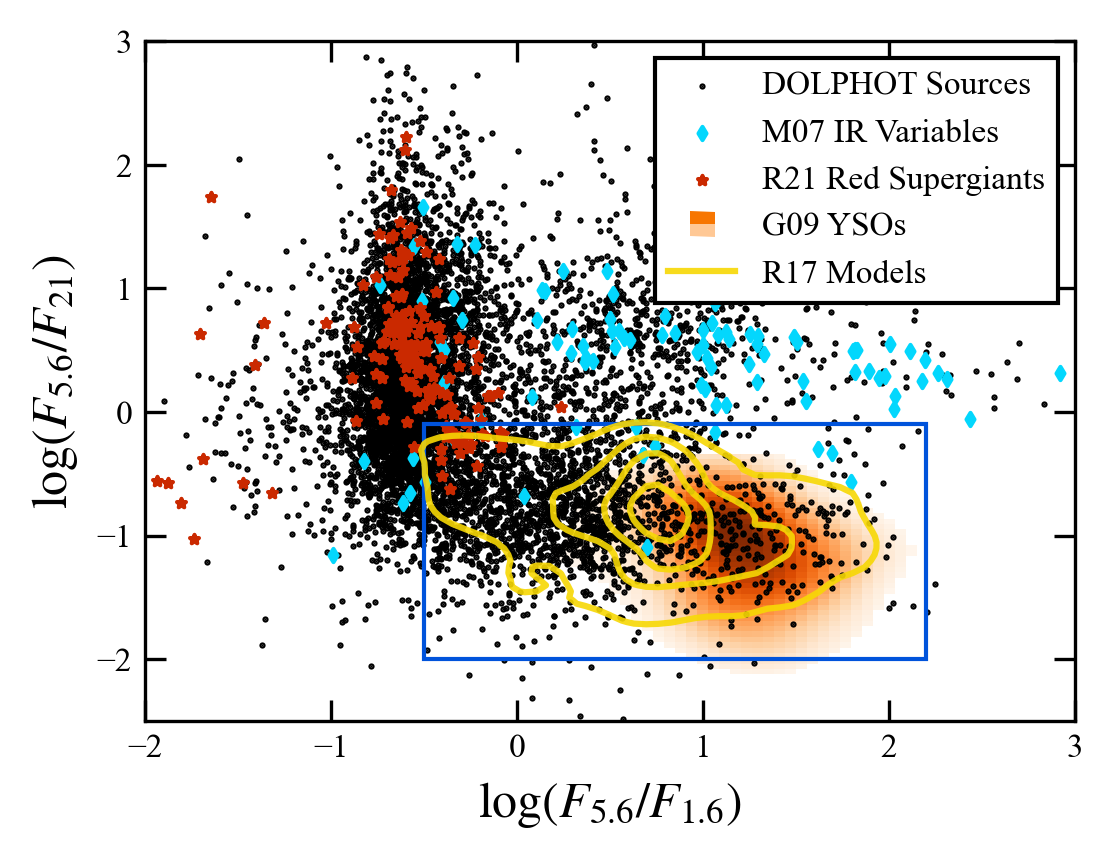}
\caption{colour-colour diagram of the identified DOLPHOT sources. The yellow contours show the density of \citetalias{2017Robitaille} YSO models. The orange heat map shows the density of the \citetalias{2009Gruendl} YSOs. Spatially cross-matched infrared variable stars from \citet{2007McQuinn} are marked as light blue diamonds. Spatially cross-matched red supergiants from \citet{2021Ren} are marked as red stars. The blue box shows our candidate YSO selection.}
    \label{fig:colourcolour}
\end{figure}

First, we use a colour-colour diagram using the $\log (F_{5.6}/F_{21})$ colour versus the $\log (F_{5.6}/F_{1.6})$ colour and compare the positions of sources to both models and previously observed extra-Galactic YSOs. To avoid undetectable low brightness YSO models, we include only the \citetalias{2017Robitaille} models that are above the 5$\sigma$ detection limits in the filters F560W and F2100W (1~$\mu$Jy and 6~$\mu$Jy at the distance of M33). This constraint leaves 10741 models with one or more viewing angles, providing 82060 unique fluxes in each filter. The colours corresponding to these YSO models are shown in Figure \ref{fig:colourcolour} as orange contours. 

For comparison, we also show the colours of 1172 LMC YSOs from \citetalias{2009Gruendl}. We use the closest available filters to match our observations, specifically, 2MASS H, IRAC 5.8, and MIPS 24 to match $F_{1.6}$, $F_{5.6}$, and $F_{21}$. We then convert the fluxes from these filters to our filters using the flux conversion derived from the simulated observation of the  \citetalias{2017Robitaille} models. The density of the YSOs \citep[Table 10 and 11;][]{2009Gruendl} are shown in Figure \ref{fig:colourcolour} as an orange heat map.

We begin by selecting only sources that cover the same colour-colour space as the \citetalias{2017Robitaille} YSO models and the LMC YSOs. This selection is shown in Figure \ref{fig:colourcolour} as a blue box defined as
\begin{equation}
    -2<\log (F_{5.6}/F_{21})<-0.1
\end{equation}
and 
\begin{equation}
    -0.6<\log (F_{5.6}/F_{1.6})<2.2.
\end{equation}
This selection yields a sample of 2248 sources. The LMC YSOs are primarily found in the redder region of this cut ($\log (F_{5.6}/F_{1.6})$ > 0.5), but there is a small population ($\approx$6\%) of LMC YSOs in the bluer region as well. Based on the density of sources and the model contours, it is likely that many of the sources we identify in the bluer region are likely contaminants. The following cuts will mostly remove sources in this bluer region. 

The stars that occupy the space above where the YSO models are located (i.e., bluer in $F_{5.6}/F_{21}$) have colours characteristic of red supergiants and AGB stars. We assess how these two contaminants populate our colour space using existing catalogues. AGB candidate stars were catalogued in M33 by \citet{2007McQuinn} using the \textit{Spitzer Space Telescope} based on their variability and infrared colours. After spatially cross-matching with our catalogue, we see that these stars cover the upper right of our colour-colour diagram. We also spatially cross-matched our DOLPHOT sources with a catalogue of red supergiants in M33 from \citet{2021Ren} obtained using the United Kingdom Infra-Red Telescope (UKIRT). The red supergiants cover the upper left of our colour-colour diagram. 

\begin{figure}
	\includegraphics[width=\columnwidth]{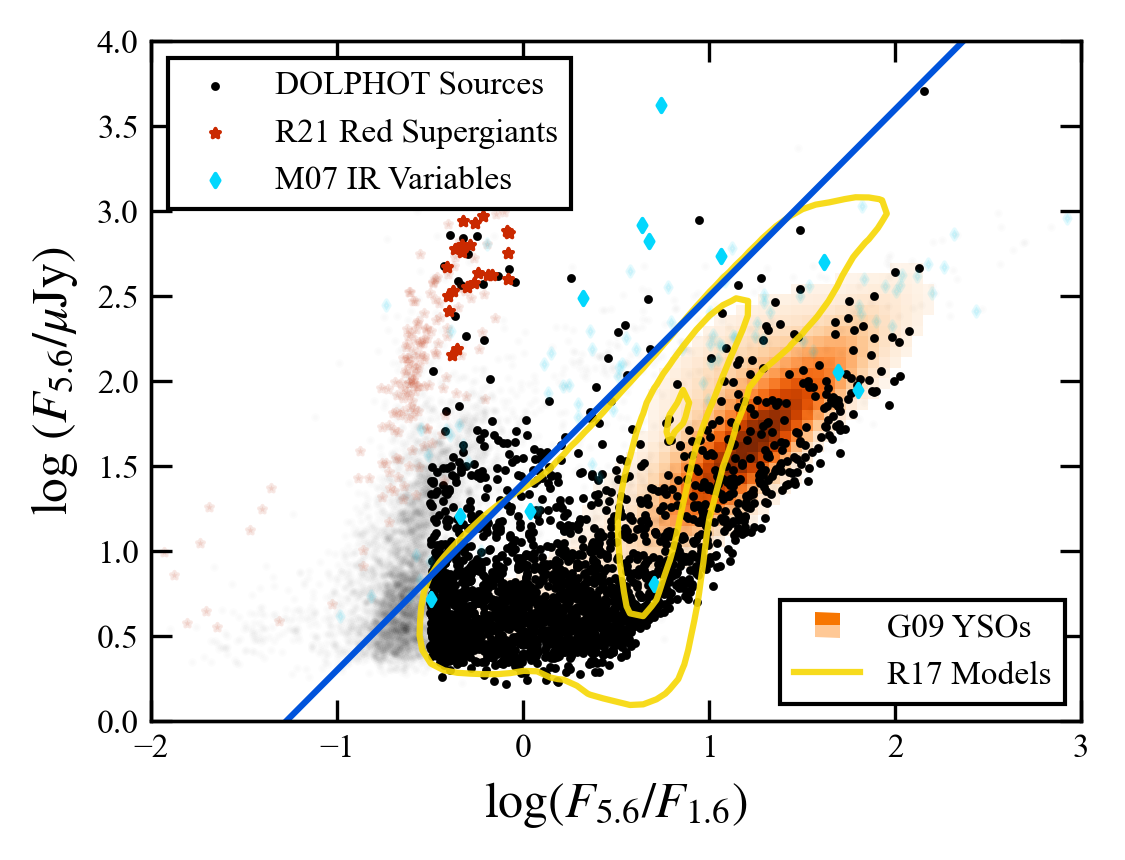}
\caption{CMD with the same sources from Figure \ref{fig:colourcolour} but focused on the sources in the blue selection box. The DOLPHOT sources, infrared variable stars \citep{2007McQuinn}, and red supergiants \citep{2021Ren} contained in the blue box are highlighted and large. All sources outside the blue box are presented as faded and small. The yellow contours show the density of \citetalias{2017Robitaille} YSO models. The orange heat map shows the density of the \citetalias{2009Gruendl} YSOs. The area below the blue line shows the additional colour cut.}
    \label{fig:cmd}
\end{figure}

We next incorporate a colour cut based on the $\log (F_{5.6})$ versus $\log (F_{5.6}/F_{1.6})$ CMD. Figure \ref{fig:cmd} shows this CMD, highlighting sources remaining from the first colour cut. We perform a second colour cut by only selecting sources that cover the same region as the YSO models and LMC YSOs in the CMD. Where the flux from the LMC YSOs has been converted to the appropriate band and scaled to the distance of M33. This second cut is defined as
\begin{equation}
    \log(F_{5.6})<1.1 \log (F_{5.6}/F_{1.6})+1.4,
\end{equation}
leaving 2005 sources.

Next, we apply a spatial cut to the remaining sources based on the overlap with GMCs as identified by CO emission. This spatial cut is necessary because background galaxies and extreme AGB stars cover a similar colour space to YSOs \citepalias{2009Gruendl}. Even with more infrared bands and visual inspection of each source, it can be challenging to remove all contaminants without a full infrared spectrum \citepalias{2009Seale}. Because YSOs should be located within GMCs, we select sources that are spatially correlated with GMCs to remove a majority of these contaminants. Of our 2005 sources, 1373 overlap spatially with a GMC. The 632 objects removed are mostly found in the bluer area of the colour-colour diagram and CMD, which is where \citetalias{2009Gruendl} found mostly background galaxies. We find that many of these 632 removed objects have the characteristic disk shape of background galaxies. Assuming all of the sources not contained in GMCs are contaminants, we can find the number of contaminants expected. Since GMCs cover $\approx$19\% of the MIRI survey area, the 632 contaminants are spread over 81\% of the survey area. Therefore, $\approx 120$ uniformly distributed contaminants are expected to overlap but be unconnected to GMCs. From the off-galaxy background observations, which should contain only background galaxies, we find 19 background galaxies that are within the colour cuts. Since the off-galaxy field is a factor of 25 smaller, we estimate there should be a total of 475 background galaxies in the survey area. Therefore, of the $\approx 120$ uniformly distributed contaminants, $\approx 90$ of them are background galaxies, with the remainder likely being extreme AGB stars.

\begin{figure}
	\includegraphics[width=\columnwidth]{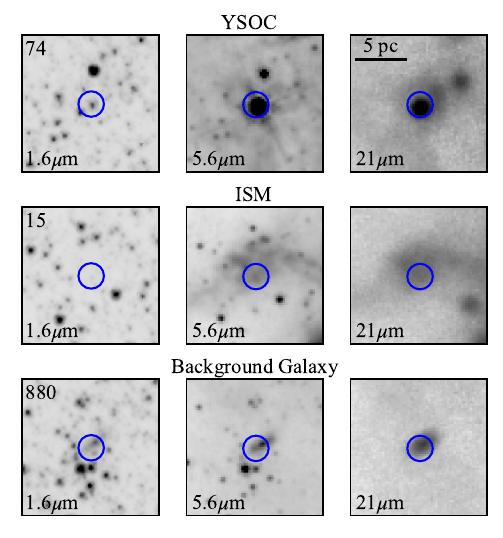}
\caption{Example of the three types of sources from visual inspection. The top, middle, and bottom row shows a YSOC, ISM, and background galaxy, respectively. The left, middle, and right shows the F160W, F560W, and F2100W filters, respectively.}
    \label{fig:exSour}
\end{figure}

Lastly, we exclude sources that are ISM material only or background galaxies that do not appear as point sources. While DOLPHOT is excellent at identifying point sources in stellar-dominated fields, the MIRI filters contain significant amounts of continuum emission. This continuum emission results in many non-point sources remaining even after all of our cuts. To remove these potential ISM sources, we manually inspect the 1373 remaining sources and eliminate any that do not appear as point sources in at least two out of our three filters. Most of the sources we removed had a dim point source in F160W but appeared diffuse and extended in the MIRI filters and were mostly found in the bluer area of the colour-colour diagram and CMD. During the manual inspection, there were 54 sources that appeared to be extended background galaxies. After removing these likely ISM and background galaxy detections, we are left with a final catalogue of 793 YSOCs, which are listed in Table \ref{tab:YSOC} and Figure \ref{fig:exSour} shows an example of each type of source. Assuming we removed all ISM detections, we are left with $\approx66$ expected contaminates ($\approx$36 background galaxies and $\approx$30 AGB stars) or $\approx8$\%, which is similar to the percentage of contaminants \citetalias{2009Seale} identified from the \citetalias{2009Gruendl} YSOs in the LMC ($\approx$6\%). 

\begin{figure}
	\includegraphics[width=\columnwidth]{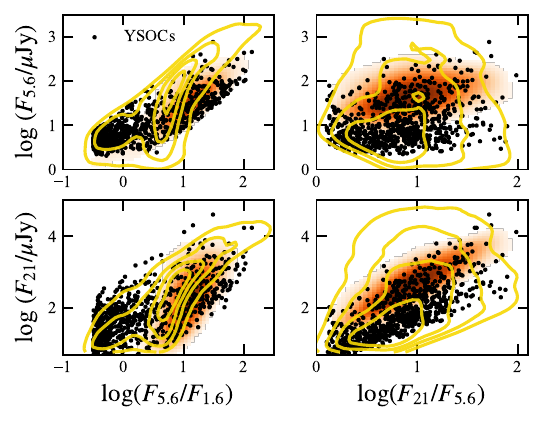}
\caption{CMDs of the final sample of YSOCs. The YSOCs are plotted on four CMDs as black points along with yellow contours showing the \citetalias{2017Robitaille} YSO models, and the orange heat map shows the density of the \citetalias{2009Gruendl}} YSOs. In all CMDs, there is a discrepancy between the model predictions and the observed YSOs from the LMC. 
    \label{fig:YSOCCMDs}
\end{figure}

Figure \ref{fig:YSOCCMDs} shows the final selection of YSOCs on four CMDs. All of these CMDs show a discrepancy between \citetalias{2017Robitaille} models and the observed \citetalias{2009Gruendl} YSOs. The \citetalias{2009Gruendl} population is dominated by bright and red sources, while the \citetalias{2017Robitaille} models are concentrated on the dimmer bluer region. This discrepancy likely comes from an observational bias to more massive YSOs since they are brighter and easier to identify. Our distribution of YSOCs in Figure \ref{fig:YSOCCMDs} shows we have captured the bright and red population traced by \citetalias{2009Gruendl} observations and the dimmer bluer population traced by \citetalias{2017Robitaille} models.

Significant detections have fluxes $F_{1.6}> 1~\mu\mathrm{Jy}$, $F_{5.6}>3~\mu\mathrm{ Jy}$, and $F_{21}>10~\mu\mathrm{Jy}$, which is broadly consistent with expectations from JWST and \textit{HST} sensitivity limit estimates. Artificial star tests will provide better estimates for how well we are eliminating ISM sources, which will be explored further in Peltonen et al. (in preparation). Given these sensitivities, we estimate that we should be sensitive to YSOs with main sequence stellar masses $M>6~M_\odot$. To derive this limit, we compared the older YSO models of \citet{2006Robitaille} to the flux limits in our three bands. The more recent \citetalias{2017Robitaille} models eschew mass estimates in favour of instantaneous radius and luminosity measurements, recognizing the ambiguities in converting a single point in a star's accretion history to its final main sequence mass. In this preliminary work, we use this mass sensitivity estimate as an approximate value to interpret our YSO count results. However, a more careful treatment connecting YSO emission models with families of accretion histories should provide better estimates of stellar mass (Richardson et al., submitted). 

Our observed number of massive YSOs is also broadly consistent with this mass sensitivity limit, given M33's star formation rate.  We estimate the number of YSOs that should be visible in the survey region using a \citet{2001Kroupa} IMF and a star formation rate in our survey area of 0.04$\pm0.01$ $M_\odot~\mathrm{yr}^{-1}$ estimated using the maps of \citet{2015boquien}. \citet{2011Mottram} split the lifetime of massive YSOs before the main sequence into $t_\mathrm{MYSO}$ and $t_\mathrm{CHII}$, which have large uncertainties. Combining these two timescales gives lifetimes of $6^{+6}_{-3}\times 10^{5}$~yr, consistent with other studies in the Milky Way \citep{2013Duarte, 2017Tige,2018Motte}. For 20 mass bins above $6~M_\odot$ that \citet{2011Mottram} use, we calculate the number of YSOs in that mass bin over their lifetime using our star formation rate. Given these estimates, we find that there should be $500^{+300}_{-200}$ YSOs with masses $M>6~M_\odot$ in the survey region, which is consistent with the 793 candidates we identified. It should be noted that the uncertainties in this estimate are significant, and our YSOC sample may be missing some real YSOs, especially those of lower masses while also containing contaminants. However, the number of uniformly distributed contaminants, the visual inspection, and the match between candidates and the expected number show that our YSOCs are likely a good representation of the massive YSOs in this region.

Finally, we note that these massive YSOCs are also likely associated with stellar clusters.  Even with JWST's excellent resolution ($< 0.4$~pc), these individual sources likely still confuse the bright sources with surrounding lower mass YSOs, and these bright sources may also be compact clusters. Future NIRCam observations of this region (e.g., JWST program GO-2130, PI: J.C. Lee) will help resolve these ambiguities, but we will proceed with the assumption that these are single medium mass YSOs.

\begin{table*}
\centering
\caption{M33 Young Stellar Object Candidates.}
 \label{tab:YSOC}
 \begin{tabular}{cccccccc}
  \hline
RA (ICRS) & DEC (ICRS) & $F_{21}$ ($\mu$Jy) & Uncertainty $F_{21}$ ($\mu$Jy) & $F_{5.6}$ ($\mu$Jy) &Uncertainty $F_{5.6}$ ($\mu$Jy)&  $F_{1.6}$ ($\mu$Jy)&Uncertainty $F_{1.6}$ ($\mu$Jy) \\
\hline
23.407269 & 30.531584 &  112.11 &   0.21 &  9.19 &   0.05 &  8.38 &   0.05 \\
23.403533 & 30.531788 &    5.02 &   0.08 &  2.67 &   0.03 &  3.96 &   0.03 \\
23.430886 & 30.529794 & 2216.29 &   0.00 & 77.70 &   0.07 &  3.57 &   0.03 \\
23.434956 & 30.535916 &  828.76 &   0.00 & 81.81 &   0.08 &  2.22 &   0.03 \\
23.436628 & 30.533236 &  302.57 &   0.28 & 41.88 &   0.08 &  1.88 &   0.03 \\
23.436573 & 30.533152 &  279.27 &   0.26 & 49.57 &   0.09 &  1.65 &   0.03 \\
23.431016 & 30.529683 &  312.77 &   0.29 & 23.31 &   0.06 &  3.70 &   0.03 \\
23.435110 & 30.535837 &  203.81 &   0.19 & 12.35 &   0.05 &  4.87 &   0.04 \\
23.435474 & 30.534151 &  159.23 &   0.15 & 23.73 &   0.07 &  2.63 &   0.03 \\
23.430964 & 30.529919 &  210.68 &   0.19 & 23.08 &   0.06 &  2.36 &   0.03 \\
  \hline
 \end{tabular}
 \\(This table is available in its entirety in a machine-readable form in the online journal.)
\end{table*}

\section{Analysis}\label{ysoana}

\subsection{Cloud-scale Star Formation Rates}
\begin{figure}
	\includegraphics[width=\columnwidth]{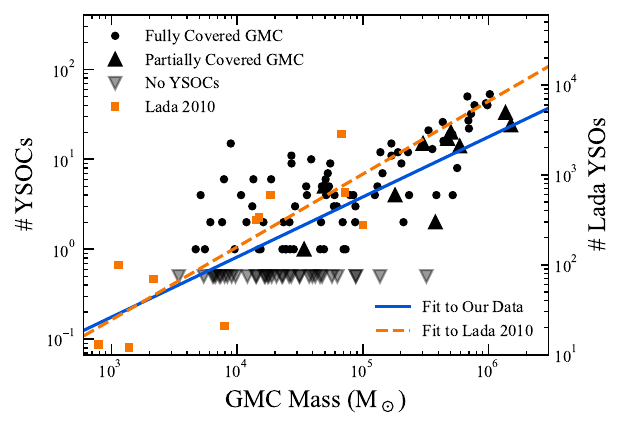}
\caption{The number of YSOCs contained within each GMC compared to the mass of that GMC. The triangles represent GMCs that are only partially contained within the MIRI survey area and are lower limits. The GMCs with no YSOCs are assigned a 0.5 and marked as upper limits. The Milky Way GMCs from \citet{2010Lada} are shown as orange squares with the total number of YSOs shown on the right axis. The right axis has been scaled by assuming a \citet{2001Kroupa} IMF with all of the YSOCs$>$6~M$_\odot$. The blue line with a slope $\alpha=0.67\pm0.06$ is from a least-squares fit to our GMCs. The orange dashed line with a slope of $\alpha=0.8\pm0.2$ is from a least-squares fit to the \citet{2010Lada} GMCs. Our GMCs show a similar correlation and scatter to Galactic GMCs.}
    \label{fig:numvmass}
\end{figure}

We now analyze the distribution of YSOCs to determine how they are affected by their cloud and galactic environment. In each GMC, we count the number of overlapping YSOCs in our catalogue and compare this number to the mass of the GMC (Figure \ref{fig:numvmass}). For GMCs only partially covered in the MIRI survey area, the number of overlapping YSOCs will be a lower limit. Clouds with no detected YSOCs are plotted with a value of 0.5 YSOCs as an upper limit in Figure \ref{fig:numvmass}. Figure \ref{fig:numvmass} shows a correlation between the number of YSOCs and GMC mass with a significant scatter.

To compare to Galactic GMCs, we use the masses and number of YSOs from \citet{2010Lada}, which includes YSOs of all masses. The YSOs from \citet{2010Lada} come primarily from \textit{Spitzer Space Telescope} observations, and the GMC masses are found from infrared extinction maps. To scale our number of high-mass YSOCs to the number of YSOs of all masses found in Milky Way GMCs, we use a \citet{2001Kroupa} IMF and integrate the total number of YSOs assuming we have all of the YSOs with final main sequence mass >$6$~M$_\odot$. 

As shown in Figure \ref{fig:numvmass}, both our data and Galactic studies find a similar correlation and scatter between the number of YSOs and GMC mass. For our data we include all limits and do a least-square fit to obtain
\begin{equation}
N_\mathrm{YSOC,>6~M_\odot} = \left(4 \pm 3 \right) \ \left(\frac{M_\mathrm{GMC}}{10^{5}~M_\odot}\right)^{0.67\pm 0.06}
\end{equation}
with a 0.4~dex spread around our fit (blue line) for both our GMCs and the \citet{2010Lada} GMCs.  

We also note that the probability of containing a YSOC rises with cloud mass, with only 50\% of clouds with $M<2\times 10^{4}~M_\odot$ hosting a YSOC but all clouds with $M>4\times 10^{5}~M_\odot$ host a YSOC.  This progression of star formation activity with increasing cloud mass is consistent with previous work in the LMC \citep[e.g.,][]{kawamura09}.

If we again assume that the number of observed YSOs we recover are tracing the number of stars with $M>6~M_\odot$ from a fully sampled Kroupa IMF, we can estimate the total mass found in YSOs for each molecular cloud. We estimate that the instantaneous ratio of the mass found in YSOs compared to cloud mass is $M_\mathrm{YSO}/M_\mathrm{GMC} \approx 5\times 10^{-3}$, similar to $\approx 7\times 10^{-3}$, found for the Solar neighbourhood using the \citet{2010Lada} GMCs. When we only consider the GMCs in our sample with $M_\mathrm{GMC}<10^5 M_{\odot}$, we get $M_\mathrm{YSO}/M_\mathrm{GMC} \approx 7\times 10^{-3}$, the same value as the \citet{2010Lada} clouds. The more massive clouds ($M_\mathrm{GMC}>10^5 M_{\odot}$) give $\approx 3\times 10^{-3}$, where the clouds of all masses show significant variation from these typical values. This variation comes directly from the variation in the number of YSOCs.

The mid-infrared flux that comes from each YSOC scales with the mass of the resulting star \citep{2012Klassen}. Here, we can measure the $F_{21}$ emitted only from the YSOCs contained within that GMC. This $F_{21}$ flux that we measure from the YSOCs accounts for $\approx 8\%$ of the total 21 $\mu$m flux in GMCs, with the remainder coming from diffuse emission or unresolved lower mass YSOs. The YSOC flux from the partially covered GMCs will again be lower limits. 

\begin{figure}
	\includegraphics[width=\columnwidth]{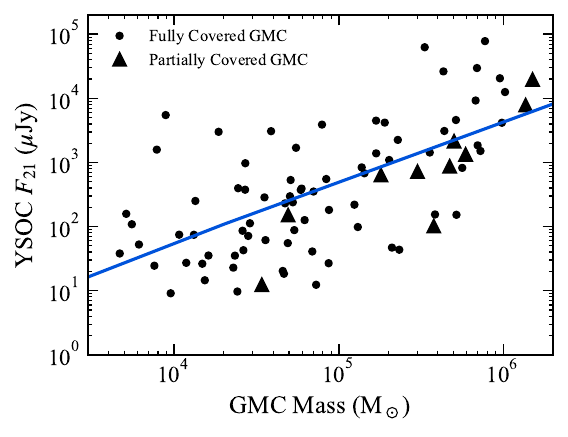}
\caption{The $F_{21}$ of the YSOCs contained within each GMC compared to the mass of that GMC. The triangles represent GMCs that are only partially contained within the MIRI survey area. We see a strong correlation between GMC mass and YSOC flux, with the blue line showing a least-squares linear fit with slope $\alpha=0.9\pm0.1$.}
    \label{fig:fluxvmass}
\end{figure}

Figure \ref{fig:fluxvmass} compares this $F_{21}$ found in YSOCs to the GMC mass. GMCs with no YSOCs are not included in this plot. Figure \ref{fig:fluxvmass} shows a clear correlation between GMC mass and the $F_{21}$ from its YSOCs. We fit a line and recover the following relation between GMC mass and $F_{21}$ flux:
\begin{equation}
    \log\left(\frac{F_\mathrm{21, YSOC}}{\mu\mathrm{Jy}}\right) =(0.9\pm0.1) \log\left(\frac{M_\mathrm{GMC}}{M_\odot}\right)-(2.0\pm0.7).
\end{equation}

 While there is a clear correlation between GMC mass and YSOC flux, there is a 0.7~dex spread around the fit. This typical scatter is larger in Figure \ref{fig:fluxvmass} than \ref{fig:numvmass}, but the scatter is more consistent across GMC mass. The scatter in the number of YSOCs is very small for high-mass GMCs and larger for low-mass GMCs. This inconsistent scatter likely comes from the stochastic nature of star formation \cite[e.g.,][]{2011Fumagalli}. For low-mass clouds, whether one or more massive stars form is fairly random and dominated by low-number statistics, but for a more massive cloud, this randomness is averaged out by the larger numbers of stars being formed. The flux coming from the YSOCs depends not only on the number but also on the mass and evolution of each YSOC \citep{2012Klassen}, which could explain the more consistent yet larger scatter seen in the flux. An additional reason for variation in YSOC number and flux could be due to GMC evolution. It is likely that some of these GMCs are earlier in their star formation evolution and have not had sufficient time to begin forming high-mass stars. 

The result from Figure \ref{fig:numvmass} and \ref{fig:fluxvmass} shows that even on cloud scales, more molecular mass results in more stars being formed. This is in contrast to star formation tracers that find a breakdown of the Kennicutt-Schmidt relation where star formation is uncorrelated with molecular gas at cloud scales,  \citep{2010Onodera,2010Schruba}, which is likely due to YSOs tracing the earlier stages of star formation better than H$\alpha$. This assertion is supported by \citet{2018Williams}, who found that when multiple star formation tracers that better trace the more embedded phase are included, the relation is preserved only with a larger scatter. Our results are largely consistent with Milky Way measurements, which find a closer correlation between molecular cloud properties and their star formation \citep{2010Lada, 2013Lada, 2021pokhrel}. This shows that by finding YSOs, we can include many phases of star formation that are not captured by a single star formation tracer alternative.

\subsection{Enhancements in Star formation from a Spiral Arm}
\label{sec:spiralarm}

We will now use our YSOCs to determine if spiral arms concentrate or enhance star formation. First, we will develop a spiral arm model based on \textsc{H~i} measurements that trace the flocculent structure seen in the young stellar populations. Then we determine how the flux coming from the YSOCs, the molecular gas content, and the ratio of these changes with distance to this model. We also compare to the grand design spiral structure seen in M33's older stellar population \citep{smercina23}, which we expect will have less of an effect on the YSOCs.

\begin{figure}
    \includegraphics[width=\columnwidth]{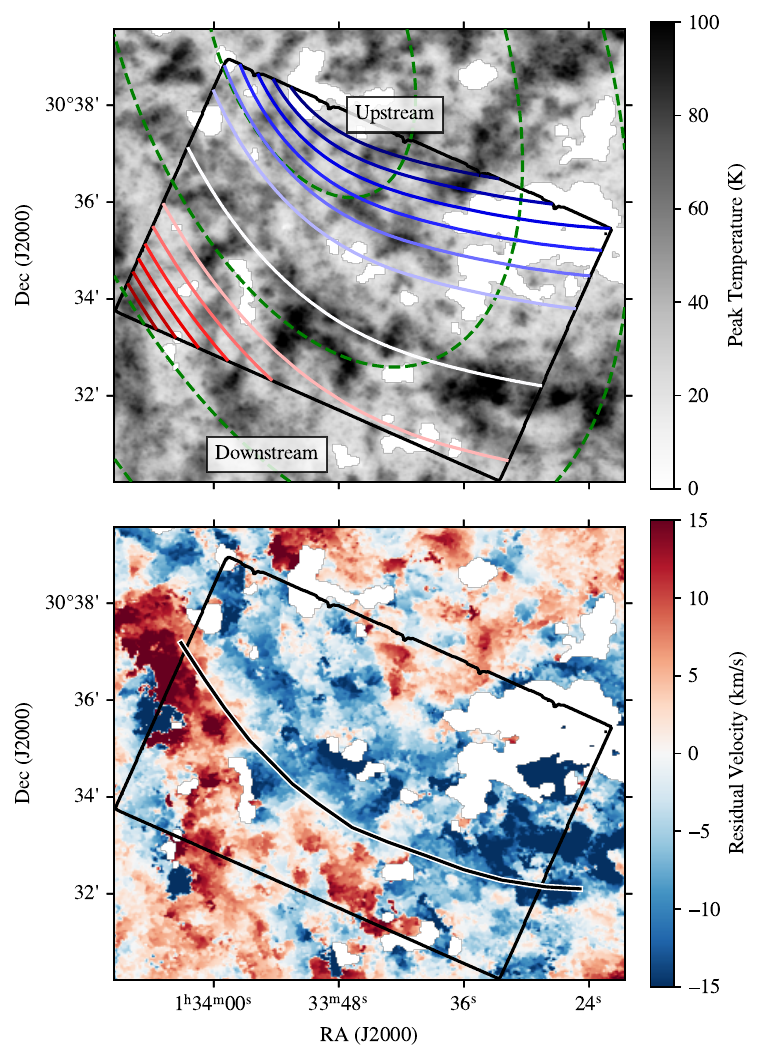}
\caption{Identification of a spiral arm segment using VLA 21-cm emission. The top map shows brightness temperature of atomic gas emission measured at each line of sight. The bottom map shows residual velocity after subtracting the projected circular velocity from the peak brightness temperature at each line of sight.
Blank regions in both maps are sightlines where no substantial \textsc{H~i} is detected, and so these regions are masked.
Both panels show the arm locus (white in top pane, black in bottom), and the box shows the extent of the JWST data. The spiral arm is chosen as the ridge of emission in the top panel associated with the blueshifted-to-redshifted transition seen in the bottom panel. In the top panel, dashed green contours show lines of constant galactocentric radius at 1, 2, and 3 kpc and the coloured lines show the distance offset from the arm ranging in 0.25 kpc intervals from $-1.5$~kpc ``upstream'' as blue to $+1.5$~kpc ``downstream'' as red.  This coordinate system is used in Figure \ref{fig:dtoarm}. Material would flow through the arm from the upper right (``upstream'') to the lower left (``downstream'').}
\label{fig:VLA21}
\end{figure}

We define the location of the spiral arm using the \textsc{H~i} 21-cm emission data from \citet{koch21}. Figure \ref{fig:VLA21} shows these data, and the black locus indicates the ridge line we define for this feature.  The top panel is the maximum brightness temperature of the emission along a given line of sight.  The bottom panel shows the velocity of that maximum emission after subtracting off the projected magnitude of the local circular velocity at that position using the rotation curve from \citet{koch18}. We identify the arm by eye as the ridge of peak brightness in the atomic gas that is coincident with the location where the cloud velocities shift from flowing into the arm region, moving faster than circular rotation (blue) to the region where they move out at a slower speed (red).  

We then fit a logarithmic spiral arm \citep{roberts75} to the locus of points identified by eye in Figure \ref{fig:VLA21}. To attain a good fit, we need to introduce a radial offset of 1.68 kpc to the start of the arm and find a best-fitting functional form of
\begin{equation}
    \phi_\mathrm{arm,g} = 0.404 \ln\left[\frac{R - 1.68~\mathrm{kpc}}{0.317~\mathrm{kpc}}\right]; R>1.70~\mathrm{kpc},
\end{equation}
where $\phi$ and $R$ are the polar angle and galactocentric radius measured in the plane of the galaxy, respectively.  Since the structure of M33 is flocculent, we also attempted to fit a hyperbolic arm model \citep{seiden79}.  However, even after including a radial offset to the hyperbolic arm model, the logarithmic spiral still provides a better fit (i.e., lower $\chi^2$).

We then define a characteristic distance with respect to the spiral arm as the minimum Cartesian distance between a point in the map and any point along the arm. The distance is assigned such that $d<0$ corresponds to ``upstream'' from the arm and $d>0$ is ``downstream.'' This measurement is only representative of the distances since material does not enter or exit spiral arms on circular orbits. Figure \ref{fig:VLA21} shows this arm distance as coloured contours (blue to red). Relating the sky position to true spatial offsets and timescales for star formation requires a flow model for material through this region, which will be presented in future work (Koch et al., in preparation).

In addition to the gas arm traced in Figure \ref{fig:VLA21}, we also consider the grand design spiral arm structure seen in the old stellar population as proposed in \citet{smercina23}.  That arm has the functional form
\begin{equation}
    \phi'_\mathrm{arm,\star} = 2.269 \ln\left(\frac{R'}{0.90~\mathrm{kpc}}\right); 0.90~\mathrm{kpc} < R' < 2.5~\mathrm{kpc}
\end{equation}
where the $\{R', \phi'\}$ coordinates are centred on a position offset with respect to the galactic centre by $\Delta \alpha=0.02^\circ$ and $\Delta \delta = 0.005^\circ$. We transform this arm into the original galactocentric frame and define distance to the arm as $d = R(\phi -\phi_\mathrm{arm,\star})$. As can be seen in Figure \ref{fig:miridat}, this older population arm is offset downstream from the star-forming gas and dust as well as the logarithmic arm. Therefore, we will proceed with primarily using the logarithmic arm defined by the \textsc{H~i} and present the stellar-derived spiral arm as a comparison.

Using the logarithmic spiral arm model defined by the gas, we determine how star formation is affected by the ISM passing through the arm. We estimate the star formation efficiency using the $F_{21}$ only coming from YSOCs per gas mass. The primary interest is how the star formation efficiency changes from upstream in the inter-arm region to downstream of the spiral arm. Therefore, to see this transition more clearly, we avoid the upstream region most distant from the arm, where there appears to be an additional flocculant concentration of material. 

\begin{figure}
	\includegraphics[width=\columnwidth]{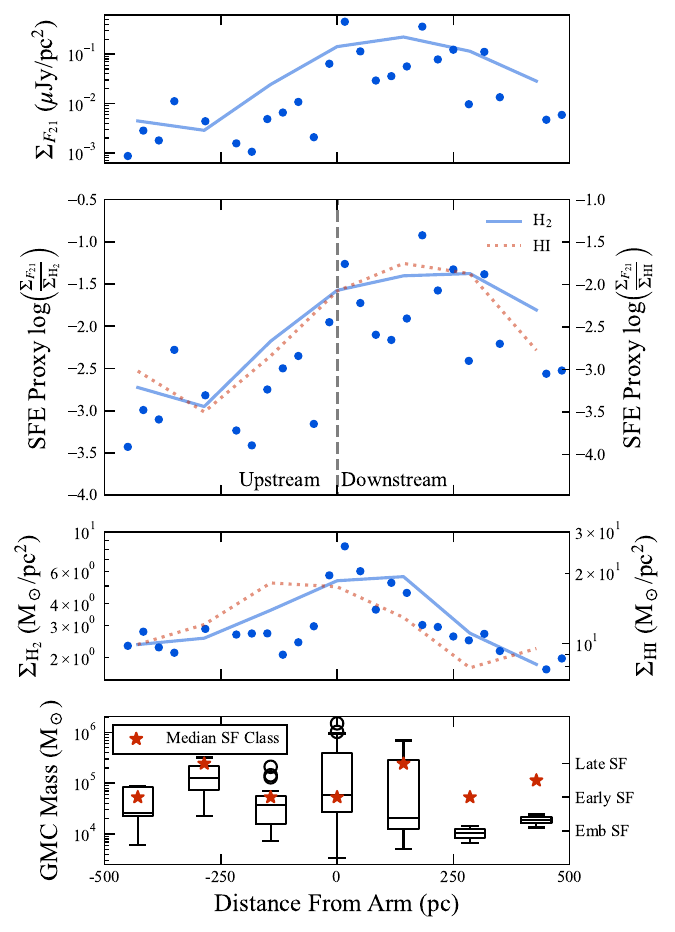}
\caption{Gas and YSOC properties with respect to the distance to the logarithmic spiral arm model. The second panel shows the estimated star formation efficiency from the total $F_{21}$ of the YSOCs normalized by the molecular mass in blue. The top and third panels show the distribution of the total $F_{21}$ of the YSOCs and the molecular mass in blue, respectively. The points show bins that are 35~pc wide, and the lines show 150~pc bins. The star formation efficiency based on \textsc{H~i} mass and the \textsc{H~i} mass is shown in red in the second and third panels with only the 150~pc bins. The bottom panel shows the properties of the GMCs in 150~pc bins, with the boxes showing the first, second, and third quartiles of GMC mass on the left axis. The error bars on the boxes show the minimum and maximum GMC mass, with outliers shown as circles. The right axis of the bottom panel shows the median star formation class identified by Koch et al. (in preparation) marked with red stars. If there are an equal number of GMCs in two star formation classes the median will be between them.}
    \label{fig:dtoarm}
\end{figure}

Figure \ref{fig:dtoarm} shows the relation between flux per gas and distance to the logarithmic spiral arm model where negative distances are upstream and positive distances are downstream from the arm. We use 35~pc and 150~pc distance bins. Figure \ref{fig:dtoarm} shows a clear peak in the molecular and atomic gas mass (third panel), where the two gas species trace a similar trend. However, the atomic gas mass rises and decreases before the molecular gas. The YSOC flux (top panel) peaks just downstream of the logarithmic model. The logarithmic arm model shows a clear peak in the star formation efficiency (second panel) corresponding to the peak in gas mass and a clear positive trend in star formation as material moves across the arm. This peak in efficiency does not closely correspond to the GMC mass or star formation class (bottom panel). Here, the `star formation class' comes from star formation tracers Koch et al. (in preparation) with four categories: no star formation, embedded star formation, early star formation, and late star formation. The logarithmic arm model shows that star formation is more efficient where there is a greater concentration of molecular and atomic gas. Overall, we see an increase in $F_{21}/M_\mathrm{H_2}$ by a factor of $>30$ from before to after the spiral arm.

We performed a similar analysis on the \citet{smercina23} arm again with the star formation efficiency from $F_{21}$ per molecular gas mass. We find that the peak in molecular gas for the arm is 1~kpc upstream of the arm, showing a clear offset between the gas and stellar content. There is no clear peak in $F_{21}$ using this spiral arm model. We find an increase in efficiency just before the \citet{smercina23} arm. However, the increase in efficiency is less pronounced than we find with the logarithmic spiral arm.

Our results are consistent with an enhancement of star formation activity in spiral arms above what would be expected from just the amount of (CO-traced) molecular or atomic gas alone. In Figure \ref{fig:dtoarm}, we see the efficiency peak just after the spiral arm, and the efficiency remains high until $\approx500$~pc after the spiral arm. Since the GMC mass does not correlate strongly with efficiency, it appears this effect does not simply come from more massive GMCs being more efficient. The median GMC mass decreases after the spiral arm, which could indicate that GMCs are being built up by the arm and then depleted by star formation moving across the arm. However, if this were the case, we would expect the star formation class to increase as GMC decreases.  

The 500-pc size scale for this enhancement of star formation efficiency is a relatively large distance in terms of the star formation process.  If the material is flowing through the arm feature with a speed of $\sim10$~km~s$^{-1}$ (e.g., Figure \ref{fig:VLA21}) and the simple geometry adopted in our arm model is appropriate, then the time to traverse the region of enhanced efficiency would be $\sim 50$~Myr.  This time is longer than the 10-20 Myr evolutionary timescales for clouds assumed globally \citep{2022Chevance} and also measured in M33 \citep[11-15~Myr]{corbelli17,Peltonen}.  We thus conclude the enhancement seen in star formation efficiency would persist over a few cloud lifetimes (or alternatively evolutionary cycles).  This conclusion depends on our simple arm offset model and an assumed speed, though more refined flow models are unlikely to change the timescales to traverse the 500 pc scale by the factor of 3-5 needed to make this timescale consistent with a single evolutionary timescale of the molecular gas.

Most studies using star formation tracers both in grand design \citep{2017Leroy,2022Williams} and flocculent spirals \citep{2010Foyle} found no significant enhancement in star formation efficiency in the spiral arms. This lack of enhancement has also been found in several simulations \citep{2020Smith,2020Kim,2021Treb}. However, \citet{2018Hirota} found that GMCs in the arm of M83 have much higher star formation efficiency than GMCs in the inter-arm region. \citet{2015Eden} conducted a YSO-based study of star formation efficiency in the Milky Way and found some enhancement in the spiral arms, but the sample was too limited to provide conclusive evidence. There are four explanations for why our results disagree with many of the results coming from star formation tracers. One, it could be that the most embedded phase of star formation, not seen by many tracers, is important to see this increase in star formation efficiency. Two, M33's flocculent spiral arms are fundamentally different from many of the more defined spiral arms. Three, the interarm region in M33 is less efficient than the interarm regions of other galaxies sampled. Four, the $F_{21}$ is enhanced due to the evolutionary state of the YSOCs in the arm, which we will explore using modelling in Peltonen et al. (in preparation).

\section{Conclusion}\label{conc}
Using new JWST MIRI observations along with near-infrared observations from PHATTER, we have constructed a large catalogue of infrared point sources in M33. These point sources were identified using the DOLPHOT JWST/MIRI module, which provides point-source photometry. We identify potential massive YSOs from these point sources using colour cuts on a colour-colour diagram and a CMD based on the \citetalias{2017Robitaille} YSO models (Figure \ref{fig:colourcolour} and \ref{fig:cmd}). The GMCs identified from the ALMA ACA CO survey allow us to remove contaminants by only including potential YSOs that are inside GMCs. Finally, we remove ISM contaminants through manual inspection. These colour cuts and contaminant removal leave 793 YSOCs. Our main findings are: 

\begin{enumerate}
    \item More massive GMCs host more YSOCs following a power-law slope of $\alpha=0.67\pm0.06$ consistent with Milky Way GMCs with a power-law slope of $\alpha=0.8\pm0.2$. The scatter around these fits is also consistent with a 0.4 dex spread for our GMCs and those in the Milky Way.
    \item More massive GMCs contain almost a directly proportional amount of YSOC flux with a power-law slope of $\alpha=0.9\pm0.1$.
    \item Star formation becomes more efficient by a factor of $>30$ across M33's flocculent spiral arm, which cannot only be attributed to an increase in GMC mass.
\end{enumerate}

In following papers, we will perform artificial star tests, which will allow us to determine the completeness of our point sources. We will also present a more complete modelling of these YSO candidates to estimate their mass and age. Equipped with these measurements of our YSOCs, we can then make more direct estimates of star formation rates in the region.

\section*{Acknowledgements}

We thank the anonymous referee for providing comments that led to improvements in quality and clarity of the paper. Based on observations with the NASA/ESA/CSA James Webb Space Telescope obtained from the Data Archive at the Space Telescope Science Institute, which is operated by the Association of Universities for Research in Astronomy, Incorporated, under NASA contract NAS5-03127. Support for program number JWST-GO-02128 was provided through a grant from the STScI under NASA contract NAS5-03127. JP and ER acknowledge support from the Natural Science and Engineering Research Council Canada, Funding Reference RGPIN-2022-03499 and from the Canadian Space Agency, Funding Reference 22JWGO1-20.  

EWK acknowledges support from the Smithsonian Institution as a Submillimeter Array (SMA) Fellow and the Natural Sciences and Engineering Research Council of Canada.

AKL gratefully acknowledges support by grants 1653300 and 2205628 from the National Science Foundation, by award JWST-GO-02128.002-A, and by a Humboldt Research Award from the Alexander von Humboldt Foundation.

AG acknowledges support from the NSF under grants AST 2008101 and CAREER 2142300.

The Flatiron Institute is funded by the Simons Foundation.

\section*{Data Availability}

The JWST MIRI and parallel NIRCam observations can be obtained from the Mikulski Archive for Space Telescopes (MAST). The PHATTER products are also available from MAST at \url{http://dx.doi.org/10.17909/t9-ksyp-na40}.

This paper makes use of the following ALMA data: ADS/JAO.ALMA\#2017.1.00901.S and 2019.1.01182.S. 
ALMA is a partnership of ESO (representing its member states), NSF (USA) and NINS (Japan), together with NRC (Canada), MOST and ASIAA (Taiwan), and KASI (Republic of Korea), in cooperation with the Republic of Chile. The Joint ALMA Observatory is operated by ESO, AUI/NRAO and NAOJ. The National Radio Astronomy Observatory is a facility of the National Science Foundation operated under cooperative agreement by Associated Universities, Inc. These data are available through the ALMA archive.

The VLA \textsc{H~i} observations are available in the NRAO archive under projects 14B-088 and 17B-162.
These data products will be publicly released in the first data release from the Local Group L-band Survey (LGLBS; E. Koch et al. in preparation).

 All other analysis data are available from the authors upon receiving a reasonable request.



\bibliographystyle{mnras}
\bibliography{example} 



\bsp	
\label{lastpage}
\end{document}